\documentstyle[prb,aps,epsfig]{revtex}
\begin{document}

\title{Momentum Distribution of a Weakly Coupled Fermi Gas}
\author{Girish S. Setlur}
\address{The Institute of Mathematical Sciences \\ Taramani, Chennai 600113,
 India.}
\maketitle

\begin{abstract}
  We apply the sea-boson method to compute
  the momentum distribution of a spinless continuum Fermi gas
  in two space dimensions
  with short-range repulsive interactions. 
  We find that the ground
  state of the system is a Landau Fermi liquid( $ 0 < Z_{F} < 1 $ ).
  We also apply this method to study the one-dimensional system 
  when the interactions
  are long-ranged gauge interactions. We map the Wigner crystal phase
  of this system.
\end{abstract}

\section{Introduction}

 In this article we apply the sea-boson method that
 is now powerful enough to yield most of
 the well-known results in one-dimension,  to solve for the momentum
 distribution of electorns in the case when the electrons are 
 in a continuum in two space dimensions with short range
 repulsive interactions. In other words, the two dimensional analog
 of the spinless Luttinger model.
 As a more nontrivial application we calculate the
 momentum distribution of a spinless Fermi system in one dimension 
 with long-range confining (gauge) interactions and show that
 the system is a Wigner crystal. The momentum distribution of this
 system exhibits some unusual features that are probably new. 

\section{The Hamiltonian}

 Here we compute the momentum distribution of the spinless Fermi gas
 with short-range repulsion in two space dimensions in an effort to
 ascertain whether or not this system describes a Landau Fermi liquid. 
 This exercise is simple and is an alternative to studying the 
 Hubbard model in 2d where the algebra is quite involved. 
 We expect the answers to qualitative questions such
 as the validity of Fermi liquid theory to be the same in both models
 since both involve very short-range interactions. We have argued
 before that whenever Fermi liquid theory breaks down, it does so
 maximally. That is, it breaks down for all values of the coupling.
 Thus it is sufficient to study the weakly coupled 2D
 system with short-range interactions where the analysis is straightforward.
 The intuition gained from this can then be transplanted 
 to the 2D Hubbard model at large $ U $, which is likely to be hard to solve
 even using sea-bosons. On the other hand, there may some qualitatively
 new physics in the case of electrons with spin. This will become clearer
 when we actually decide to study the Hubbard model directly
 in future publications.
 In the present case, the hamiltonian in the sea-boson language is given by,
\begin{equation}
H = \sum_{ {\bf{k}} {\bf{q}} } \left( \frac{ {\bf{k.q}} }{m} \right)
A^{\dagger}_{ {\bf{k}} }({\bf{q}})A_{ {\bf{k}} }({\bf{q}})
 + \sum_{ {\bf{q}} \neq 0} \frac{ v_{ {\bf{q}} } }{2V}
\sum_{ {\bf{k}} {\bf{k}}^{'} }
[A_{ {\bf{k}} }(-{\bf{q}}) + A^{\dagger}_{ {\bf{k}} }({\bf{q}})]
[A_{ {\bf{k}}^{'} }({\bf{q}}) + A^{\dagger}_{ {\bf{k}}^{'} }(-{\bf{q}})]
\end{equation}
 This hamiltonian describes a self-interacting Fermi gas provided we 
 assume that $ v_{ {\bf{q}} } = \frac{ v_{0} }{m}\mbox{     } \theta(\Lambda - |{\bf{q}}|) $
 where $ \Lambda \ll k_{F} $ here $ 0 < v_{0} \ll 1 $ is a dimensionless 
 parameter. We may solve for the boson occupation number as follows.
\begin{equation}
\left< A^{\dagger}_{ {\bf{k}} }({\bf{q}})A_{ {\bf{k}} }({\bf{q}}) \right>
 = \frac{1}{V}
\sum_{i} \frac{ \Lambda_{ {\bf{k}} }(-{\bf{q}}) }
{ (\omega_{i} + \frac{ {\bf{k.q}} }{m})^{2} }g^{2}_{i}(-{\bf{q}})
\end{equation}
where $ \Lambda_{ {\bf{k}} }(-{\bf{q}}) = n_{F}({\bf{k}}-{\bf{q}}/2)(1 - n_{F}({\bf{k}}+{\bf{q}}/2)) $.
\begin{equation}
g^{-2}_{i}(-{\bf{q}}) = \frac{1}{V} \mbox{    }\sum_{ {\bf{k}} }
 \frac{ n_{F}({\bf{k}}-{\bf{q}}/2) - n_{F}({\bf{k}}+{\bf{q}}/2) }
{ (\omega_{i} - \frac{ {\bf{k.q}} }{m})^{2} }
\end{equation}
\begin{equation}
\epsilon_{RPA}({\bf{q}},\omega) = 
1 + \frac{ v_{ {\bf{q}} } }{V} \sum_{ {\bf{k}} }
 \frac{ n_{F}({\bf{k}}+{\bf{q}}/2) - n_{F}({\bf{k}}-{\bf{q}}/2) }
{ \omega - \frac{ {\bf{k.q}} }{m} }
\end{equation}
 As argued in earlier works, we have to interpret sum over modes with
 care so as to not lose the particle-hole mode, the collective mode
 being obvious. The is particularly important in two dimensions where 
 we expect both to be present. Thus the sum over modes is defined as follows.
\begin{equation}
\sum_{i} \mbox{  }f({\bf{q}},\omega_{i})
  = \frac{ \int^{ \infty }_{0} d \omega \mbox{      }W({\bf{q}},\omega)
 \mbox{      } 
f({\bf{q}},\omega) }
{  \int^{ \infty }_{0} d \omega \mbox{      }W({\bf{q}},\omega) }
\end{equation}
Here the weight function is given by,
\begin{equation}
W({\bf{q}},\omega) = -Im \left( \frac{1}{ \epsilon_{RPA}
({\bf{q}},\omega - i0^{+}) } 
 \right)
\end{equation}
 In our earlier work, we had suggested that in the above formula we
 have to use a dielectric function that is sensitive to
 significant qualitative changes in one-particle properties. 
 The simple RPA dielectric function does not possess qualities
 we expect from a Wigner crystal. Thus we shall have to derive a
 new dielectric function using the localised basis rather than the
 plane-wave basis. In the case of the Fermi gas in two dimensions
 we find that the system is a Landau Fermi liquid and there is no
 need to use a better dielectric function, the simple RPA suffices.
 The momentum distirbution is always given by,
\begin{equation}
< n_{ {\bf{k}} } > = \frac{1}{2} 
 \left[ 1 + e^{ -2S^{0}_{B}({\bf{k}}) } \right] \mbox{          }
n_{F}({\bf{k}}) 
 + \frac{1}{2} 
 \left[ 1 - e^{ -2S^{0}_{A}({\bf{k}}) } \right] \mbox{          }
(1-n_{F}({\bf{k}}))
\label{NBAR}
\end{equation}
\begin{equation}
S^{0}_{A}({\bf{k}}) = \sum_{ {\bf{q}} }
\left< A^{\dagger}_{ {\bf{k}}-{\bf{q}}/2 }({\bf{q}})
A_{ {\bf{k}}-{\bf{q}}/2 }({\bf{q}}) \right>
\end{equation}
\begin{equation}
S^{0}_{B}({\bf{k}}) = \sum_{ {\bf{q}} }
\left< A^{\dagger}_{ {\bf{k}}+{\bf{q}}/2 }({\bf{q}})
A_{ {\bf{k}}+{\bf{q}}/2 }({\bf{q}}) \right>
\end{equation}
The computation of the boson occupation number
 $ \left< A^{\dagger}_{ {\bf{k}} }({\bf{q}})
A_{ {\bf{k}} }({\bf{q}}) \right> $ is the key to evaluating one-particle
 properties. 

\section{ The Computations }
 
 In this section, we compute the various quantities defined
 in the previous sections. 
 In the case of a of a Fermi
 gas in two dimensions with short-range repulsive interactions we
 may use the simple RPA dielectric function. 
 The integrals are somewhat complicated
 since in two dimensions, the angular parts are very
 troublesome unlike in three dimensions. 
 Therefore we take the easy way
 out and copy the results first derived by Stern\cite{Stern}.
\begin{equation}
\epsilon^{r}_{RPA}(q,\omega) = 1 + \frac{ m k_{F} v_{q} }{2 \pi q }
\{ \frac{q}{ k_{F} } - C_{-} \left[  \left( \frac{q}{2k_{F}}
 -  \frac{ m \omega }{ k_{F} q }   \right)^2 - 1 \right]^{\frac{1}{2}}
 -  C_{+} \left[  \left( \frac{q}{2k_{F}}
 +  \frac{ m \omega }{ k_{F} q }  \right)^2 - 1 \right]^{\frac{1}{2}} \}
\end{equation}
\begin{equation}
\epsilon^{i}_{RPA}(q,\omega) = \frac{ m k_{F} v_{q} }{ 2 \pi q }
\{ D_{-} \left[ 1 - \left[ \frac{q}{2k_{F}} - \frac{ m \omega }{ k_{F} q }
 \right]^2
 \right]^{\frac{1}{2}}
- D_{+} \left[ 1 - \left[ \frac{q}{2k_{F}} + \frac{ m \omega }{ k_{F} q } 
 \right]^2
 \right]^{\frac{1}{2}} \}
\end{equation}
where,
\begin{equation}
C_{ \pm }  =  sgn \left[ \frac{q}{2k_{F}} \pm \frac{ m \omega }{ k_{F} q } \right],
\mbox{         }
D_{ \pm }  = 0, \mbox{       }
 \left|  \frac{q}{2k_{F}} \pm \frac{ m \omega }{ k_{F} q } \right| > 1 
\end{equation}
\begin{equation}
C_{ \pm }  = 0, \mbox{       } D_{ \pm } = 1, \mbox{       } 
 \left|  \frac{q}{2k_{F}} \pm \frac{ m \omega }{ k_{F} q } \right| < 1 
\end{equation}
\begin{equation}
g^{-2}(-{\bf{q}},\omega) =   \frac{ m k_{F} }{ 2 \pi q } \mbox{     }
 \mbox{     }
\{ \frac{ m C_{-} }{ k_{F} q } 
 \left[  \left( \frac{q}{2k_{F}}
 -  \frac{ m \omega }{ k_{F} q }  \right)^2 - 1 \right]^{-\frac{1}{2}}
 \left( \frac{q}{2k_{F}}
 -  \frac{ m \omega }{ k_{F} q }  \right) 
 -  \frac{ m C_{+} }{ k_{F} q } \left[  \left( \frac{q}{2k_{F}}
 +  \frac{ m \omega }{ k_{F} q } \right)^2 - 1 \right]^{-\frac{1}{2}}
 \left( \frac{q}{2k_{F}} +  \frac{ m \omega }{ k_{F} q }  \right) \}
\end{equation}
\[
\left< A^{\dagger}_{ {\bf{k}} }({\bf{q}})A_{ {\bf{k}} }({\bf{q}}) \right>
 = \frac{1}{V} \frac{1}{ Z(q) }\int^{\infty}_{0} d \omega \mbox{       }W(q,\omega) 
 \mbox{       } \frac{ \Lambda_{ {\bf{k}} }(-{\bf{q}}) }
{ (\omega + \frac{ {\bf{k.q}} }{m})^{2} } \mbox{      } 
 g^{2}(-{\bf{q}},\omega)
\]
\begin{equation}
W(q,\omega) = \frac{ \epsilon^{i}_{RPA}(q,\omega) }
{ \epsilon^{r 2}_{RPA}(q,\omega) 
 +  \epsilon^{i 2}_{RPA}(q,\omega) }
\end{equation}
\begin{equation}
Z(q) = \int^{ \infty }_{0} d \omega \mbox{        }W(q,\omega) 
\end{equation}
In general we have,
\begin{equation}
S^{0}_{A}({\bf{k}}) = \frac{1}{ (2 \pi)^2 } 
\int^{\infty}_{0} dq \mbox{        }q \mbox{     }
\frac{1}{ Z(q) }\int^{\infty}_{0} d \omega \mbox{       }W(q,\omega) 
 \mbox{       } f_{A}(k,q,\omega) 
\mbox{      } 
 g^{2}(-{\bf{q}},\omega)
\end{equation}
\begin{equation}
S^{0}_{B}({\bf{k}}) = \frac{1}{ (2 \pi)^2 } 
\int^{\infty}_{0} dq \mbox{        }q \mbox{     }
\frac{1}{ Z(q) }\int^{\infty}_{0} d \omega \mbox{       }W(q,\omega) 
\mbox{      } 
f_{B}(k,q,\omega)
 g^{2}(-{\bf{q}},\omega)
\end{equation}
\begin{equation}
f_{A}(k,q,\omega) = \int^{2 \pi}_{0} d\theta 
\mbox{       }\frac{ \theta( k_{F}^2 - k^2 - q^2 + 2 k q cos(\theta) ) }
{ ( \omega + \frac{ kq }{m} cos(\theta) - \epsilon_{q} )^2 }
\end{equation}
\begin{equation}
f_{B}(k,q,\omega) = \int^{2 \pi}_{0} d\theta 
\mbox{       }\frac{ \theta( k^2 + q^2 -k_{F}^2 + 2 k q cos(\theta) ) }
{ ( \omega + \frac{ kq }{m} cos(\theta) + \epsilon_{q} )^2 }
\end{equation}
Define,
\begin{equation}
u(x;A,B) \equiv \int dx \mbox{       }\frac{ 1 }{ (A + B \mbox{    }Cos[x])^2 }
\end{equation}
From $ Mathematica^{TM} $ we find,
\[
u(x;A,B) = - 2 A \frac{
\mbox{      }ArcTanh \left[ \frac{ (A-B) Tan \left[ \frac{x}{2} \right] }
{ \sqrt{ B^2 - A^2 } } \right] }
{ (A^2 - B^2) \sqrt{ B^2 - A^2 } }
\]
\begin{equation}
 + \frac{ B \mbox{       } Sin[x] } 
{ (B^2 - A^2) (A + B \mbox{   }Cos[x]) }
\end{equation}
Integrating by parts we find,
\begin{equation}
f_{A}(k,q,\omega)
 = \int^{2 \pi}_{0} d\theta \mbox{       }sin \theta \mbox{      }
\mbox{      }\delta( \frac{ k_{F}^2 - k^2 - q^2 }{ 2 k q } + cos(\theta) ) 
\mbox{      }u(\theta ; \omega - \epsilon_{q}, \frac{ k q }{m})
\end{equation}
\begin{equation}
f_{B}(k,q,\omega)
 = \int^{2 \pi}_{0} d\theta  \mbox{       }sin \theta \mbox{      }
\mbox{       }\delta( \frac{ k^2 + q^2 -k_{F}^2 }{ 2 k q } + cos(\theta) ) 
\mbox{       }u(\theta ; \omega + \epsilon_{q}, \frac{ k q }{m})
\end{equation}
This may be rewritten as,
\[
f_{A}(k,q,\omega)
 = \int^{\pi}_{0} d\theta \mbox{       }sin \theta \mbox{      }
\mbox{       } \delta( \frac{ k_{F}^2 - k^2 - q^2 }{ 2 k q } + cos(\theta) ) 
\mbox{     }u(\theta ; \omega - \epsilon_{q}, \frac{ k q }{m})
\]
\begin{equation}
 - \int^{\pi}_{0} d\theta \mbox{       }sin \theta \mbox{      }
\mbox{       } \delta( \frac{ k_{F}^2 - k^2 - q^2 }{ 2 k q } - cos(\theta) ) 
\mbox{     }u(\theta + \pi ; \omega - \epsilon_{q}, \frac{ k q }{m})
\end{equation}
\[
f_{B}(k,q,\omega)
 = \int^{\pi}_{0} d\theta  \mbox{       }sin \theta \mbox{      }
\mbox{       }\delta( \frac{ k^2 + q^2 -k_{F}^2 }{ 2 k q } + cos(\theta) ) 
\mbox{        }u(\theta ; \omega + \epsilon_{q}, \frac{ k q }{m})
\]
\begin{equation}
 - \int^{\pi}_{0} d\theta  \mbox{       }sin \theta \mbox{      }
\mbox{       }\delta( \frac{ k^2 + q^2 -k_{F}^2 }{ 2 k q } - cos(\theta) ) 
\mbox{        }u(\theta + \pi ; \omega + \epsilon_{q}, \frac{ k q }{m})
\end{equation}

\begin{equation}
\theta_{0} = arccos \left[ \frac{ k^2 + q^2 - k_{F}^2 }{2 k q } \right]
\end{equation}
\begin{equation}
\theta^{'}_{0} = arccos \left[ \frac{ - k^2 - q^2 + k_{F}^2 }{2 k q } \right]
\end{equation}
\[
f_{A}(k,q,\omega) = 
 u(\theta_{0} ; \omega - \epsilon_{q}, \frac{ k q }{m})
\left[  \theta( \frac{ k_{F}^2 - k^2 - q^2 }{ 2 k q } + 1 ) 
 -  \theta( \frac{ k_{F}^2 - k^2 - q^2 }{ 2 k q } - 1 )  \right]
\]
\begin{equation}
 - u(\theta^{'}_{0} + \pi ; \omega - \epsilon_{q}, \frac{ k q }{m}) 
 \left[ \theta( 1 - \frac{ k_{F}^2 - k^2 - q^2 }{ 2 k q } ) 
 -  \theta( - 1 - \frac{ k_{F}^2 - k^2 - q^2 }{ 2 k q } )  \right]
\end{equation}
\[
f_{B}(k,q,\omega) = 
 u(\theta^{'}_{0} ; \omega + \epsilon_{q}, \frac{ k q }{m})
\left[ \theta( \frac{ k^2 + q^2 -k_{F}^2 }{ 2 k q } + 1 )
 -  \theta( \frac{ k^2 + q^2 -k_{F}^2 }{ 2 k q } - 1 ) \right]
\]
\begin{equation}
 - u(\theta_{0} + \pi ; \omega + \epsilon_{q}, \frac{ k q }{m})
\mbox{       }
\left[ \theta( 1 - \frac{ k^2 + q^2 -k_{F}^2 }{ 2 k q } ) 
 - \theta( -1 - \frac{ k^2 + q^2 -k_{F}^2 }{ 2 k q } )  \right]
\end{equation}
 The collective mode occurs when $ Im[ \epsilon ] = 0 $, that is, for small
 enough $ q $. This means that we have to treat this separately. 
\[
S^{0}_{A}({\bf{k}}) = \frac{1}{ (2 \pi)^2 } 
\int^{\infty}_{0} dq \mbox{        }q \mbox{     }
\frac{1}{ Z(q) }\int^{\infty}_{0} d \omega \mbox{       }W(q,\omega) 
 \mbox{       } f_{A}(k,q,\omega) 
\mbox{      } 
 g^{2}(-{\bf{q}},\omega)
\]
\begin{equation}
 + \frac{1}{ (2 \pi)^2 } 
\int^{\infty}_{0} dq \mbox{        }q \mbox{     } 
 \mbox{       } f_{A}(k,q,\omega_{c}) 
\mbox{      } 
 g^{2}(-{\bf{q}},\omega_{c})
\end{equation}
\[
S^{0}_{B}({\bf{k}}) = \frac{1}{ (2 \pi)^2 } 
\int^{\infty}_{0} dq \mbox{        }q \mbox{     }
\frac{1}{ Z(q) }\int^{\infty}_{0} d \omega \mbox{       }W(q,\omega) 
\mbox{      } 
f_{B}(k,q,\omega)
 g^{2}(-{\bf{q}},\omega)
\]
\begin{equation}
+ \frac{1}{ (2 \pi)^2 } 
\int^{\infty}_{0} dq \mbox{        }q \mbox{     }
f_{B}(k,q,\omega_{c})
 g^{2}(-{\bf{q}},\omega_{c})
\end{equation}
 Here it is implicit that in $ W $ we assume that $ Im[ \epsilon ]  \neq 0 $.
 The dispersion of the collective mode may be found using
$ {\it{ Mathematica }}^{TM} $. It is given below.
\begin{equation}
\omega_{c}(q) = q \mbox{  }(2 \pi + m v_{q}) 
\frac{ (\pi^2 q^2 + m \pi q^2 v_{q} + k_{F}^2 m^2 v_{q}^2)^{\frac{1}{2}} }
{ 2 m^2 \sqrt{ \pi } v_{q} \sqrt{ \pi + m v_{q} } }
\end{equation}
 This dispersion is real and positive for all $ q $ and for all $ v_{q} > 0 $.
 Thus in the small $ q $ limit, where using just the RPA dielectric
 function is justified and is also the limit where
 the close to the Fermi surface features of the momentum distribution is
 given exactly, we are justified in retaining only the coherent
 part. Thus we may write,
\begin{equation}
S^{0}_{A}({\bf{k}}) \approx \frac{1}{ (2 \pi)^2 } 
\int^{\infty}_{0} dq \mbox{        }q \mbox{     } 
 \mbox{       } f_{A}(k,q,\omega_{c}) 
\mbox{      } 
 g^{2}(-{\bf{q}},\omega_{c})
\end{equation}
\begin{equation}
S^{0}_{B}({\bf{k}}) \approx \frac{1}{ (2 \pi)^2 } 
\int^{\infty}_{0} dq \mbox{        }q \mbox{     }
f_{B}(k,q,\omega_{c})
 g^{2}(-{\bf{q}},\omega_{c})
\end{equation}
 To determine whether or not Fermi liquid theory breaks down,
 we have to compute,
\begin{equation}
S^{0}_{A}(k_{F}) \approx \frac{1}{ (2 \pi)^2 } 
\int^{\infty}_{0} dq \mbox{        }q \mbox{     } 
 \mbox{       } f_{A}(k_{F},q,\omega_{c}) 
\mbox{      } 
 g^{2}(-{\bf{q}},\omega_{c})
\label{EQNSAF}
\end{equation}
\begin{equation}
S^{0}_{B}(k_{F}) \approx \frac{1}{ (2 \pi)^2 } 
\int^{\infty}_{0} dq \mbox{        }q \mbox{     }
f_{B}(k_{F},q,\omega_{c})
 g^{2}(-{\bf{q}},\omega_{c})
\label{EQNSBF}
\end{equation}
 If $ S^{0}_{A}(k_{F}), S^{0}_{B}(k_{F}) < \infty $ then the ground state
 is a Landau Fermi liquid.
 If $  S^{0}_{A}(k_{F}) = S^{0}_{B}(k_{F}) = \infty $ then the system
 is a non-Fermi liquid. For small $ q $ if we set
 $ \omega_{c} = v_{eff} \mbox{      }q $ we have,
\begin{equation}
f_{A}(k_{F},q,\omega_{c}) \sim f_{B}(k_{F},q,\omega_{c}) \sim 1/q^2
\end{equation}
Also,
\begin{equation}
g^2(-{\bf{q}},\omega_{c}) \sim q
\end{equation}
 Thus the integrals in Eq.(~\ref{EQNSAF}) and Eq.(~\ref{EQNSBF}) are 
 infrared finite. This means that
 $ S^{0}_{A}(k_{F}), S^{0}_{B}(k_{F}) < \infty $ and the system is a Landau
 Fermi liquid. The details of the momentum distribution can be worked out but
 are not terribly important.

\section{One Dimensional System with Long-Range Interactions}

 In this case, we expect the system to
 be a Wigner crystal. Thus we have to be careful about
 the choice of the dielectric function.
 First, we postulate that $ v_{ q } = 2e^2/(q\mbox{  }a)^2 $
 which corresponds to the gauge potential. Here $ a $ has
 dimensions of length and $ e^2 > 0 $ is dimensionless. 
 From the form of this potential, one hopes that
 we need not concern ourselves with the issues that
 were relevant in the case of the Calogero-Sutherland model namely the
 repulsion attraction duality (more prosaically called
 back-scattering). Thus we may write
 as before,
 \[
\left< A^{\dagger}_{ {\bf{k}} }({\bf{q}})A_{ {\bf{k}} }({\bf{q}}) \right>
 = \frac{1}{V} \frac{1}{ Z(q) }\int^{\infty}_{0} d \omega
 \mbox{       }W(q,\omega) 
 \mbox{       } \frac{ \Lambda_{ {\bf{k}} }(-{\bf{q}}) }
{ (\omega + \frac{ {\bf{k.q}} }{m})^{2} } \mbox{      } 
 g^{2}(-{\bf{q}},\omega)
\]
\begin{equation}
W(q,\omega) = \frac{ \epsilon^{i}(q,\omega) }
{ \epsilon^{r 2}(q,\omega) 
 +  \epsilon^{i 2}(q,\omega) }
\end{equation}
\begin{equation}
Z(q) = \int^{ \infty }_{0} d \omega \mbox{        }W(q,\omega) 
\end{equation}
\begin{equation}
 g^{-2}(-{\bf{q}},\omega) = \frac{1}{ v_{q} }
\frac{ \partial }{ \partial \omega } \epsilon({\bf{q}},\omega)
\end{equation}
  As mentioned before,
  we have to be extra careful in making sure that we choose the
 right dielectric function. The RPA-dielectric function is not likely
 to suffice since its static structure factor (SSF) does not exhibit the 
 features we expect from a Wigner crystal. In particular, we expect
 $ S(2k_{F}) = \infty $ as we shall see soon. 
 To convince ourselves of this we ascertain the properties of the 
 RPA dielectric function with long-range interactions.
\begin{equation}
\epsilon^{r}_{RPA}(q,\omega) = 1 + v_{q} \frac{m}{2 \pi q}
Log \left[ \frac{ (k_{F}+q/2)^2 - \left(\frac{m \omega }{q}\right)^2 }
{ (k_{F}-q/2)^2 - \left( \frac{ m \omega }{q} \right)^2 } \right]
\end{equation}
The zero of the above dielectric function gives us the dispersion
 of the collective modes.
\begin{equation}
\omega_{c}(q) = \frac{ |q| }{m} 
\sqrt{ \frac{ (k_{F}+q/2)^2 - (k_{F}-q/2)^2 exp(- \frac{ 2 \pi q }{m v_{q}} ) }{ 1 -  exp(- \frac{ 2 \pi q }{m v_{q} }) } }
\end{equation}
 For $ |q| \ll k_{F} $ and $ v_{q} = 2 e^2/ (a q)^2 $ we find, 
\begin{equation}
\omega_{c}(q) \approx \frac{1}{m} \sqrt{ \frac{ e^2 k_{F} m }{a^2 } }
\sqrt{ \frac{2}{ \pi } }
 + \frac{ a^2 k_{F} \sqrt{ \frac{ e^2 k_{F} m }{ a^2 } }
 \sqrt{ \frac{ \pi }{2} } q^2 }{ 2 e^2 m^2 } + O(q^4)  
\end{equation}
 This plasmon-like gap
 $ \omega_{0} \equiv \frac{1}{m} \sqrt{ \frac{ e^2 k_{F} m }{a^2 } }
\sqrt{ \frac{2}{ \pi } } $ in the collective mode is present due to the
 characteristic $ 1/q^2 $ nature of the potential. But this is also
 present in the three dimensional electron gas and is not a sign of
 an insulator since the latter is not at high densities. A gap in
 the {\it{ one-particle }} Green function at the Fermi momentum could be taken
 as a sign of insulating behaviour\cite{Sen}. However, in our approach we
 are unable to compute the full Green function as yet. Thus we must resort
 to a more indirect approach.
 For a Wigner crystal, the SSF must
 exhibit certain singularities.  
 Thus we have to use the generalised-RPA
 that is sensitive to qualitative changes
 in single-particle properties. The new dielectric function will
 involve the full momentum distribution which has to be determined 
 self-consistently using the above sea-boson equations.
 In our earlier work we suggested that the new dielectric
 function should also
 involve fluctutations in the momentum distribution,
 however it now appears that
 that is fortunately not needed. The number-number
 correlation function is vanishingly small
 in the thermodynamic limit as shown in another preprint
 and this means we may simply write,
\begin{equation}
\epsilon({\bf{q}},\omega) = 1 + \frac{v_{q}}{L}
\sum_{ k } \frac{ {\bar{n}}_{k+q/2} - {\bar{n}}_{k-q/2} }
{ \omega - \xi_{k+q/2} + \xi_{k-q/2} }
\label{WIGDI}
\end{equation}
\begin{equation}
\xi_{k} = \frac{ k^2 }{2m} - \sum_{ q \neq 0 }
 \frac{ v_{q} }{L} {\bar{n}}_{k-q}
\end{equation}
 and the momentum distribution is determined self-consistently
 using the sea-boson equations (Eq.(~\ref{NBAR})).
 This is too difficult to solve analytically and hence we have to resort to
 a numerical solution.
 In order to simplify proceedings even further, we use only the collective
 mode. The particle-hole mode which is due to a nonzero $ Im[\epsilon] $
 is needed if one is interested in features of the momentum distribution away
 from the Fermi surface more accurately. However we shall hope that 
 this is given not too badly even at these regions far from the
 Fermi points. 
\begin{equation}
\left< A^{\dagger}_{ {\bf{k}} }({\bf{q}})A_{ {\bf{k}} }({\bf{q}}) \right>
 = \frac{1}{V} 
 \mbox{       } \frac{ \Lambda_{ {\bf{k}} }(-{\bf{q}}) }
{ (\omega_{c}(q) + \frac{ k.q }{m})^{2} } \mbox{      } 
 g^{2}(-{\bf{q}},\omega_{c})
\end{equation}
\begin{equation}
g^{-2}(-{\bf{q}},\omega_{c}) = \frac{ m \omega_{c} }{ \pi q }
\left[ \frac{1}{ \omega_{c}^2 - (v_{F}q + \epsilon_{q})^2 }
- \frac{1}{ \omega_{c}^2 - (v_{F}q - \epsilon_{q})^2 } \right]
\end{equation}
\begin{equation}
S^{0}_{A}(k) = \frac{1}{L}\sum_{q} 
 \mbox{       } \frac{ n_{F}(k-q) }
{ (\omega_{c}(q) + \frac{ k.q }{m} - \epsilon_{q})^{2} } \mbox{      } 
 g^{2}(-{\bf{q}},\omega_{c})
\end{equation}
\begin{equation}
S^{0}_{B}(k) = \frac{1}{L}\sum_{q} 
 \mbox{       } \frac{ (1-n_{F}(k+q)) }
{ (\omega_{c}(q) + \frac{ k.q }{m} + \epsilon_{q})^{2} } \mbox{      } 
 g^{2}(-{\bf{q}},\omega_{c})
\end{equation}
 To proceed further, we have to ascertain the nature of the collective modes 
 $ \omega_{c} $. If we use the RPA-dielectric function, we find
 a constant dispersion (plasmon) 
 for small $ |q| $. 
 However we have found that this
 choice is inconsistent since if we use the momentum distribution obtained
 from this to solve for the dielectric function and recompute 
 the collective mode we obtain a completely different answer namely :
 $ \omega_{c}(q) = v_{s} |q| $.
 Therefore it is critical that we get the 
 dispersion right. It appears then that we have to use the form given in the
 appendix which is not easy to simplify.
  A systematic approach for obtaining the dispersion
  of the collective modes
  has been suggested by Sen and Baskaran\cite{Baskaran}.
  Since the plane-wave basis is not appropriate for deriving a formula 
  for the dielectric function of a Wigner crystal,
  we shall follow this approach.
  First, we would like ascertain the lattice structure in the small $ a $
  limit. In this limit, the potential energy dominates
  over the kinetic energy. If we assume that the electrons are all on
  a circle of perimeter $ L $ then to minimise the potential energy,
  we have to maximize the separation. This leads to an equally spaced
  set of lattice points with lattice constant $ l_{c} $ such that
  $ N \mbox{       }l_{c} = L $.
  Thus we have $ l_{c} = 1/\rho_{0} = \pi/k_{F} $.
  Thus we assume that the electrons all lie on a circle with equal
  spacing between them. Therefore we expect the structure factor to diverge
 for a momentum $ q_{0} = 2 \pi/l_{c} = 2k_{F} $. From the Bijl-Feynman
 formula $ S(q) = \epsilon_{q}/\omega_{c}(q) $ we may suspect that
 a choice of $ \omega_{c} $ that vanishes at $ q = 2k_{F} $ is 
 needed. The form of the dispersion is given in the appendix.
 For $ \pi/N \ll |q \mbox{   }l_{c}| \ll 2 \pi  $ it seems that
 $ \omega_{q} \approx \omega_{0} $.
 For $ |q \mbox{   }l_{c}| \ll \pi/N  $ we have to be more careful.
 And of course we must have $ \omega_{q} = 0 $ for
 $ q \mbox{   }l_{c} = \pm 2 \pi $.
 But since in the thermodynamic limit $ \pi/N \approx 0 $ we may
 choose (hopefully) $ \omega_{q} \approx \omega_{0} $.
 In Fig. 1 and 2 we see the momentum distribution obtained from these
 formulas has been plotted. In fact, we may write down a closed formula
 for the momentum distribution. 
\begin{equation}
{\bar{n}}_{k} = \frac{1}{2} 
( 1 +  Exp \left[  -\frac{ m \mbox{  }\omega_{0} }
{ k_{F}^2 - k^2 } \right] ) n_{F}(k)
 +  \frac{1}{2} 
( 1 - Exp \left[ -\frac{ m \mbox{  }\omega_{0} }
{ k^2 - k_{F}^2 } \right] ) (1-n_{F}(k))
\end{equation}
 The striking feature of this momentum distribution is that it is perfectly
 flat at $ |k| = k_{F} $. In other words, not only is the slope zero but
 all the derivatives of the momentum distribution vanish at $ |k| = k_{F} $. 
 This is a striking prediction.
 This may be contrasted with the smooth Gaussian function
 of Gori-Giorgi and Ziesche \cite{Gori}( Eq.(B1) in their Appendix B ).
 But they consider three dimensional systems which may be different from
 the one studied here.
 One particle spectral functions
 are accessible to tunneling experiments or angle-resolved
 photoemission spectroscopy(ARPES).
 A more difficult problem may be to
 experimentally realise a 1d electron system with long-range gauge
 interactions. 


\begin{figure}[h]
\centerline{\psfig{file=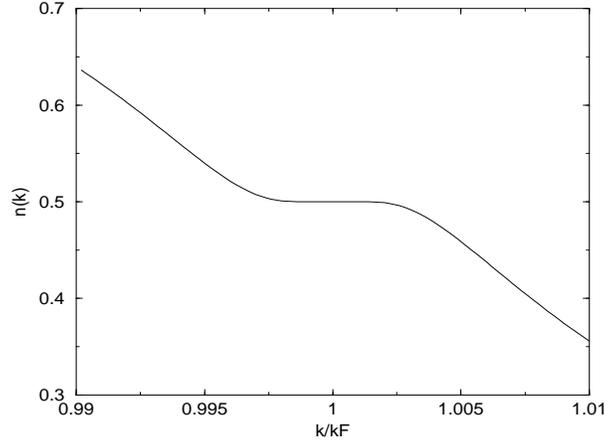,width = 8cm,height=6cm}}
\vspace*{8pt}
\caption { \label{wigner1} Momentum Distribution of a
 Wigner Crystal (with zoom) } 
\end{figure}

\begin{figure}[h]
\centerline{\psfig{file=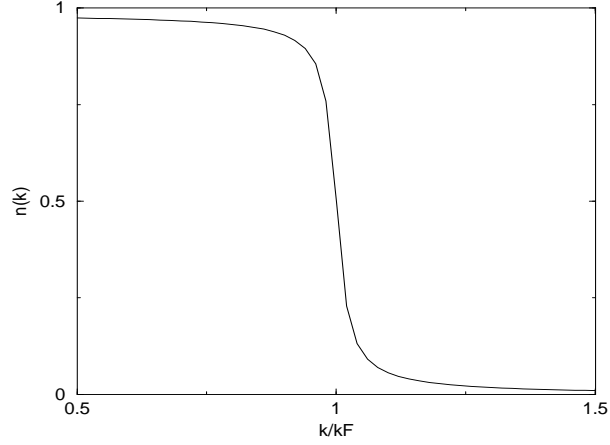,width = 8cm,height=6cm}}
\vspace*{8pt}
\caption{ \label{wigner2} Momentum Distribution of a Wigner Crystal (no zoom) }
\end{figure}
 The formula below for the static structure factor 
  is derived in the appendix.
\begin{equation}
S(q) =  \frac{1}{N} \mbox{   }
 \frac{ sin^2( \frac{ q N \mbox{   }l_{c} }{2} ) }
{  sin^2( \frac{ q \mbox{   }l_{c} }{2} )  }  \mbox{         }
e^{ - \frac{ q^2 }{ m \omega_{0} }  }
\end{equation}
This may be further simplified in the thermodynamic limit as follows.
Consider,
\begin{equation}
\delta(x) \approx \frac{ sin ( N \mbox{   }x ) }{ \pi \mbox{   }x }
\end{equation}
Then we may write,
\begin{equation}
S(q) =   \delta(q) \frac{ 2 \pi }{ l_{c} }
+  \delta(q-2k_{F}) \frac{ 2 \pi }{ l_{c} }
  \mbox{         }
e^{ - \frac{ 4 k_{F}^2 }{ m \omega_{0} }  }
\end{equation}
 As we can see here the strucutre factor
 diverges at $ |q| = 2k_{F} $ which means
 the system is a Wigner crystal ( a true Wigner crystal, since
 the divergence is from a delta-function )

\section{Appendix}

 Here we use the approach suggested by Sen\cite{Baskaran} to derive a formula
 for the collective modes.
 The formula Eq.(~\ref{WIGDI}) although probably right is not very
 illuminating, for it is hard to see how the structure factor derived from
 this formula possesses the features we expect namely a divergence
 at $ |q| = 2k_{F} $. Thus we would like to derive a formula for 
 the dielectric function where this feature is manifest. To do this we
 adopt the localised basis rather than the plane-wave basis.
 In real space, the hamiltonian we are studying is written as follows.
\begin{equation}
H = \sum_{i=0}^{N-1} \frac{ p^2_{i} }{2m} - \frac{ e^2 }{ a^2 } \sum_{i > j }
|x_{i}-x_{j}|
\label{HAM}
\end{equation}
 We assume that particles are on a circle and $ |x| $ is the chord length. 
 We would like to compute the dielectric function using this model.
 The density operator in momentum space is,
\begin{equation}
\rho_{ {\bf{q}} } = \sum_{m = 0}^{N-1}e^{i q \left( m \mbox{   }l_{c}
 + {\tilde{x}}_{m} \right) }
\end{equation}
 Here $ x_{m} = m \mbox{   }l_{c} + {\tilde{x}}_{m} $
 is measured along the circumference of the circle
 (see Fig. 3 below).
\begin{figure}[h]
\centerline{\psfig{file=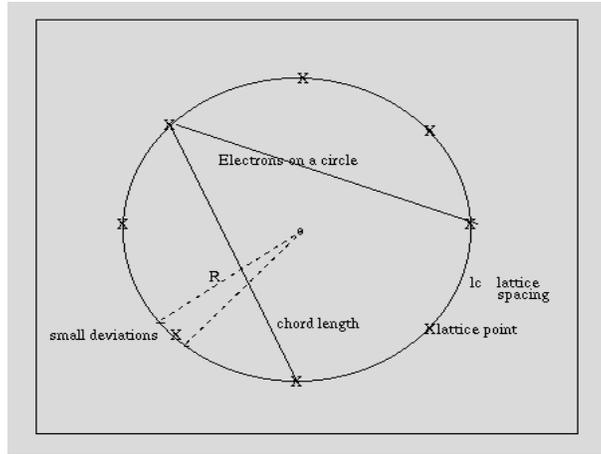,width = 8cm,height=6cm}} 
\vspace*{8pt}
\caption { \label{schema} Schematic Diagram of Electrons on a 
 Circular Lattice }
\end{figure}
 The average density is given by,
\begin{equation}
\left< \rho_{ {\bf{q}} } \right>  
 = \sum_{m = 0}^{N-1}
e^{ i q \mbox{  } m \mbox{   }l_{c}  }
e^{- \frac{1}{2} q^2 \mbox{  }  < {\tilde{x}}^2_{m} > }
\end{equation}
It can be shown that (see below)
 $   < {\tilde{x}}^2_{n} > = 1/m \omega_{0} $, independent of 
the index $ n $.
Thus we have,
\begin{equation}
\left< \rho_{ {\bf{q}} } \right>  
 = \frac{1 - e^{ i q N \mbox{   }l_{c}  } }
{ 1 -  e^{ i q  \mbox{   }l_{c}  } }
e^{ - \frac{ q^2 }{ 2 m \omega_{0} }  }
\end{equation}
 In those instances where $ <\rho_{q}> \neq 0 $ the static structure factor
 is given by,
\begin{equation}
S(q) \equiv \frac{ | <\rho_{ {\bf{q}} }> |^2 }{N} = 
  \frac{1}{N} \mbox{   }
\left( \vline \frac{1 - e^{ i q N \mbox{   }l_{c}  } }
{ 1 -  e^{ i q  \mbox{   }l_{c}  } } \vline \right)^2 \mbox{       }
e^{ - \frac{ q^2 }{ m \omega_{0} }  }
 =  \frac{1}{N} \mbox{   }
 \frac{ sin^2( \frac{ q N \mbox{   }l_{c} }{2} ) }
{  sin^2( \frac{ q \mbox{   }l_{c} }{2} )  }  \mbox{         }
e^{ - \frac{ q^2 }{ m \omega_{0} }  }
\end{equation}
 The rest of the details are as follows. We write
 $ {\tilde{x}}_{m}(t) \approx R \mbox{       }{\tilde{\theta}}_{m}(t) $. 
 In terms of the small angles $ {\tilde{\theta}}_{i} $ the hamiltonian
 in Eq.(~\ref{HAM}) may be written as follows. 
\begin{equation}
H = - \frac{1}{2m R^2} \sum_{m=0}^{N-1} \mbox{   }
\frac{ \partial^2 }{ \partial {\tilde{ \theta }}^2_{m} }
 - \frac{ R \mbox{      }e^2 }{ 2a^2 } \sum_{m \neq m^{'} }
\left[ ( cos ( 2 \pi \frac{ m }{N} + {\tilde{\theta}}_{m} )
 - cos ( 2 \pi \frac{ m^{'} }{N} + {\tilde{\theta}}_{ m^{'} } ) )^2
 + ( sin ( 2 \pi \frac{ m }{N}  + {\tilde{\theta}}_{m}  )
 - sin (2 \pi  \frac{ m^{'} }{N} + {\tilde{\theta}}_{ m^{'} }   ) )^2
 \right]^{\frac{1}{2}}
\end{equation}
 We may expand the above hamiltonian in powers of the angle and retain only
 the leading terms to arrive at the following hamiltonian in the harmonic
 approximation. 
\begin{equation}
 H =  \sum_{n=0}^{N-1} \mbox{   }
\frac{ {\tilde{p}}_{n}^2 }{2m} 
 + \sum_{ n \neq n^{'} }
  A(n,n^{'}) \mbox{         }
( {\tilde{x}}_{n} -  {\tilde{x}}_{ n^{'} })^2
 +  \sum_{ n \neq n^{'} } B(n,n^{'}) \mbox{          }
 ( {\tilde{x}}_{n} -  {\tilde{x}}_{ n^{'} }) 
\end{equation}
where,
\begin{equation}
 A(n,n^{'}) = \frac{ \pi e^2 }{ 4 L a^2 } \mbox{         }
\vline sin( \pi \frac{ (n-n^{'}) }{N} ) \vline
\end{equation}
\begin{equation}
 B(n,n^{'}) = - \frac{ e^2 }{2a^2} 
\mbox{       }sgn( sin( \pi \frac{ (n-n^{'}) }{N} ) )
\mbox{        }cos ( \pi \frac{ (n-n^{'}) }{N} )
\end{equation}
 Despite appearences to the contrary, the extremum of the potential 
 is at $ {\tilde{x}}_{n} \equiv 0 $. Since $ A > 0 $, this extremum is
 also a minimum. One has to now  compute the various correlation
 function of the system. The primary one
 of interest is,
\begin{equation}
 G_{11}(nt;n^{'}t^{'}) = <T  \mbox{  } {\tilde{x}}_{n}(t) \mbox{  } {\tilde{x}}_{n^{'}}(t^{'}) > 
\end{equation}
The other is,
\begin{equation}
 G_{21}(nt;n^{'}t^{'}) =
 <T  \mbox{  } {\tilde{p}}_{n}(t) \mbox{  } {\tilde{x}}_{n^{'}}(t^{'}) > 
\end{equation}
Thus we have,
\begin{equation}
i \frac{ \partial }{ \partial t }  G_{11}(nt;n^{'}t^{'})
 = \frac{i}{m}  G_{21}(nt;n^{'}t^{'})
\end{equation}
\begin{equation}
i \frac{ \partial }{ \partial t }  G_{21}(nt;n^{'}t^{'})
 = \delta_{n, n^{'} } \delta(t-t^{'})
 -4i \sum_{ j \neq n }
A(n,j) ( G_{11}(nt;n^{'}t^{'}) - G_{11}(jt;n^{'}t^{'}) ) 
\end{equation}
This may be solved by a Fourier transform.
\begin{equation}
G_{ij}(nt;n^{'}t^{'}) = \frac{1}{ -i \beta } \sum_{p}
e^{ z_{p} ( t-t^{'} ) } \frac{1}{N} \sum_{q} 
 e^{ i q \mbox{   } l_{c} ( n-n^{'}) } {\tilde{G}}_{ij}(q,z_{p})
\end{equation}
Thus we have,
\begin{equation}
i \mbox{   }z_{p} \mbox{         }  {\tilde{G}}_{11}(q,z_{p})
 = \frac{i}{m}  {\tilde{G}}_{21}(q,z_{p})
\end{equation}
\begin{equation}
i \mbox{   }z_{p}\mbox{   }{\tilde{G}}_{21}(q,z_{p})
 = 1  + 4i \mbox{      }({\tilde{A}}(q)  -{\tilde{A}}(0))
 \mbox{     }{\tilde{G}}_{11}(q,z_{p}) 
\end{equation}
\begin{equation}
{\tilde{A}}(q) = \sum_{j} A(j) e^{ i q \mbox{   }l_{c} \mbox{   }j }
\end{equation}
\begin{equation}
{\tilde{G}}_{11}(q,z_{p}) =
\left( i\mbox{   } m \mbox{   }z_{p}^2  
 -  4i \mbox{      }({\tilde{A}}(q)  -{\tilde{A}}(0)) \right)^{-1}
\end{equation}
Thus,
\begin{equation}
G_{11}(nt;n^{'}t^{'}) = \frac{1}{ N } \sum_{q} 
 e^{ i q \mbox{   } l_{c} ( n-n^{'}) }
\frac{1}{ \beta } \sum_{p}
 e^{ z_{p} ( t-t^{'} ) } 
\left( m \mbox{   }z_{p}^2  
 -  4\mbox{      }({\tilde{A}}(q)  -{\tilde{A}}(0)) \right)^{-1}
\end{equation}
\begin{equation}
{\tilde{A}}(q) = \frac{c_{0}}{2i}
\sum_{j=0}^{N-1} \left( e^{ i ( \frac{ \pi }{N} + q \mbox{   }l_{c})  j }
 -  e^{ i ( -\frac{ \pi }{N} + q \mbox{   }l_{c}) j } \right) 
 = \frac{c_{0}}{2i}
 \left( \frac{ 1 + e^{ i q \mbox{   }l_{c}  N } }
{ 1 - e^{ i ( \frac{ \pi }{N} + q \mbox{   }l_{c}) } }
 -    \frac{ 1 + e^{ i q \mbox{   }l_{c} N  } }
{ 1 - e^{ i ( -\frac{ \pi }{N} + q \mbox{   }l_{c}) } } \right) 
\end{equation}
\begin{equation}
{\tilde{A}}(0) =  \frac{c_{0}}{2i}
 \left( \frac{ 2 }
{ 1 - e^{ i \frac{ \pi }{N} } }
 -    \frac{ 2 }
{ 1 - e^{ -i \frac{ \pi }{N} } } \right)  
 \approx \frac{ k_{F} e^2 }{ 2 \pi  a^2 } 
\end{equation}
 Here $ c_{0} = \pi e^2 / 4 L a^2 $. The dispersion of the collective mode 
 is then given by,
\begin{equation}
\omega_{q} = \sqrt{ \frac{ 4 }{ m } }
\left( {\tilde{A}}(0) - {\tilde{A}}(q) \right)^{\frac{1}{2}}
\label{DISP}
\end{equation}
 If $ q = 0 $ or $ q = 2 k_{F} $ then $ \omega_{q} = 0 $.
 For a more thorough analysis one has to compute the full dielectric function
 from the DDDCF and use it to compute the full momentum distribution that
 is accurate even away from the Fermi surface. However we shall be content at
 features close to the Fermi surface.
 For the static structure factor we have to compute the equal time version 
 of the correlation function.  We find that $ \omega_{q} $ is
 in general, complex. This means that the eigenmodes also have a
 finite lifetime. Thus we  have,
\[
G_{11}(nt;n^{'}t) = \frac{1}{ N } \sum_{q} 
 e^{ i q \mbox{   } l_{c} ( n-n^{'}) }
\frac{1}{2 \pi m} \int_{-\infty}^{\infty} d z_{p} \mbox{         } 
\left( z_{p}^2  
 + \omega_{c}^{2}(q) \right)^{-1}
\]
\begin{equation}
 = \frac{1}{ 2m \mbox{  }N } \sum_{q} 
 \frac{ e^{ i q \mbox{   } l_{c} ( n-n^{'}) } }{ \omega_{c}(q) }
\end{equation}
From Eq.(~\ref{DISP}) it is clear that for
 $ \pi/N \ll |q \mbox{  }l_{c}| < 2 \pi $ we have
 $ \omega_{q} \approx \omega_{0} $ since $ {\tilde{A}}(q) \approx 0 $
 for $ q $ in this region. Since in the thermodynamic limit, this
 is all of $ q $, we shall boldy write,
\begin{equation}
< x_{n}(t) \mbox{  }x_{ n^{'} }(t) > 
  = \frac{1}{ 2m \mbox{  }N } \sum_{q} 
 \frac{ e^{ i q \mbox{   } l_{c} ( n-n^{'}) } }{ \omega_{0} }
 = \delta_{ n,n^{'} }
 \mbox{      } 
 \frac{ 1 }{ m \omega_{0} }
\end{equation}

\section{Acknowledgements}

 The author would like to thank Dr. Debanand Sa for sharing his
 extensive knowledge of Many Body Theory and
 for teaching the author how to plot the figures in this article.
 Also Mr. Akbar Jaffari's help with $ Mathematica^{TM} $ 
 is gratefully acknowledged.

\end{document}